\documentclass{emulateapj}
\usepackage{graphicx}
\shortauthors{West \& Basri}
\shorttitle{Rotation in Late-type M dwarfs}
\begin{document}

\title{A First Look at Rotation in Inactive Late-Type M Dwarfs}

\author{Andrew A. West\altaffilmark{1,2,3}, 
Gibor Basri\altaffilmark{2}}

\altaffiltext{1}{Corresponding author: aaw@space.mit.edu}
\altaffiltext{2}{Astronomy Department, University of California, 601
  Campbell Hall, Berkeley, CA 94720-3411}
\altaffiltext{3}{MIT Kavli Institute for Astrophysics and Space Research, 77
  Massachusetts Ave, 37-582c, Cambridge, MA 02139-4307}

\begin{abstract} 
  We have examined the relationship between rotation and activity in
  14 late-type (M6-M7) M dwarfs, using high resolution spectra taken
  at the W.M. Keck Observatory and flux-calibrated spectra from the
  Sloan Digital Sky Survey. Most were selected to be inactive at a
  spectral type where strong H$\alpha$ emission is quite common. We
  used the cross-correlation technique to quantify the rotational
  broadening; six of the stars in our sample have $v$sin$i$$\geq$3.5
  km\,s$^{-1}$.  Our most significant and perplexing result is that
  three of these stars do not exhibit H$\alpha$ emission, despite
  rotating at velocities where previous work has observed strong
  levels of magnetic field and stellar activity.  Our results suggest
  that rotation and activity in late-type M dwarfs may not always be
  linked, and open several additional possibilities including a
  rotationally-dependent activity threshold, or a possible dependence
  on stellar parameters of the Rossby number at which
  magnetic/activity ``saturation'' takes place in fully convective
  stars.
\end{abstract}


\section{Introduction}

Many M dwarfs, which are the most abundant stars in the Milky Way,
have strong magnetic dynamos that give rise to chromospheric and
coronal heating, producing emission from the X-ray to the radio.
Although this magnetic heating (or activity) has been observed for
decades, the exact mechanisms that control magnetic activity in M
dwarfs are still not well-understood.  In the Sun, magnetic fields are
generated at the boundary between the convective and the radiative
zones (known as the tachocline), where the differential rotation in
the convective zone creates a rotational shear
\citep{Parker93,Ossendrijver03,Thompson03} that allows magnetic
fields to be generated, stored and ultimately rise to the surface.
These fields drive the heating of the stellar chromosphere and corona,
resulting in both flares and lower level quiescent magnetic
activity. A common diagnostic of this heating in optical spectra (of M
dwarfs) is the H$\alpha$ emission line.

Rotation in solar-type stars slows with time due to angular momentum
loss from magnetized stellar winds; as a result, magnetic activity
decreases.  \citet{Skumanich72} found that both activity (as measured by
Ca II emission) and projected rotational velocity decrease over time as a power law
(t$^{-0.5}$).  Subsequent studies confirmed the Skumanich results and
demonstrated that there is a strong link between age, rotation and
activity in Solar-type stars \citep{Barry88,Soderblom91,Pizzolato03, MH08}.

There is strong evidence that the rotation-activity relation extends
from stars more massive than the Sun to smaller dwarfs
\citep{Pizzolato03, MB03, Kiraga07}.  However, at a spectral type of
$\sim$M3 \citep[0.35 M$_{\odot}$;][]{NLDS,
  Chabrier1997}\footnote{Although mass and spectral type are closely linked
  in evolved low-mass dwarfs, young, low-mass objects cool rapidly
  with age and migrate across a range of spectral types while at the
  same stellar mass; mass and spectral type are only coupled in stars
  that have settled on the main sequence ($>$ 500 Myr for most
  late-type M dwarfs).}, stars become fully convective and the tachocline
presumably disappears.  This transition marks an important change in
the stellar interior that has been thought to affect the production
and storage of internal magnetic fields. Despite this change in the
stellar interiors, magnetic activity persists in the late-type M
dwarfs; the fraction of active stars peaks around a spectral type of
M7 before decreasing into the brown dwarf regime \citep{Hawley96,
  Gizis2000, W04}.

A few previous studies of H$\alpha$ emission have uncovered evidence of a possible rotation-activity relation extending past the M3 convective transition
and into the brown dwarf regime \citep{D98, MB03, RB07}.  In addition,
recent simulations of magnetic dynamo generation in fully convective
stars find that rotation may play a role in magnetic field generation
\citep{Dobler06,Browning08}.  However, the lack
of an unbiased sample of high resolution spectra of late-type M dwarfs
complicates the situation.

Using over 30000 spectra from the Sloan Digital Sky Survey
\citep[SDSS;][]{DR6}, \citet{W06,W08} showed that the activity
fraction of M dwarfs varies as a function of stellar age (using
Galactic height as a proxy for age) and that the H$\alpha$ activity
lifetime for M6-M7.5 stars is 7-8 Gyr.  Nearby samples of late-type M
dwarfs are therefore biased towards young populations with high levels
of activity; until recently every known M7 dwarf was observed to be
magnetically active \citep{Hawley96, Gizis2000, W04}.

Due to the intrinsic faintness of late-type M dwarfs, previous
rotation studies were limited (and therefore biased) toward, bright,
nearby, and therefore active M dwarfs; the 21 M6-M7.5 dwarfs that have
previous rotation measurements are all active \citep{D98, MB03, RB07}.
The advent of large spectroscopic surveys of M dwarfs (e.g.\,SDSS)
allows for improved sample selection that can minimize observational
bias in the study of late-type M dwarfs.

In this paper we present results from our study of the $v$sin$i$
rotation velocities for a small sample of M6-M7 dwarfs, most of which
were selected to be inactive or weakly active from the SDSS low-mass
star spectroscopic sample.  We describe the sample and observations in
\S2, our analysis in \S3, and the observational results in \S4.  We
then discuss our results in \S5.

\section{Data}

Our sample was selected from the \citet{W08} Sloan Digital Sky Survey
(SDSS) M dwarf catalog, a spectroscopic sample of almost 40000 M and
L-type dwarfs.  We selected the brightest M6 and M7 stars which were
either inactive (H$\alpha$ equivalent width $<$ 0.3) or weakly active
(H$\alpha$ equivalent width $<$ 1.0).  12 stars were selected using
these criteria.  Two additional active (H$\alpha$ equivalent width
$>$ 1.0) M7 dwarfs were added to the sample for comparison to previous
studies.  While our sample is not a complete unbiased sample,
representative of the underlying M dwarf population, it does consist
of late-type M dwarfs with activity properties selectively different
than previously observed.

The 14 stars were observed during four observing runs (13 October
2006, 14 October 2006, 25 April 2007, and 24 January 2008) using the
HIRES spectrograph at Keck I.  Our setup covered the wavelength region
from 5600 to 10000 \AA~with a resolution of R$\sim$30000.  The data
were reduced with standard IDL routines that included flat-fielding,
sky subtraction and cosmic ray removal.  In addition to our 14 program
stars, we observed the nearby M6 dwarf Gl406, a star with no
measurable rotation \citep{D98, MB03} that we used as a zero rotation
velocity template.

\section{Analysis}
\subsection{Rotation from Cross Correlation}

To measure the $v$sin$i$ rotation velocities for our sample, we used a
cross-correlation technique similar to that of previous studies
\citep[e.g.\,][]{D98, MB03}: we
cross-correlated each program spectrum with the spectrum of a slowly
rotating comparison star. The width of the resulting cross correlation
function is a direct probe of the rotational broadening.

\begin{figure*}
\plotone{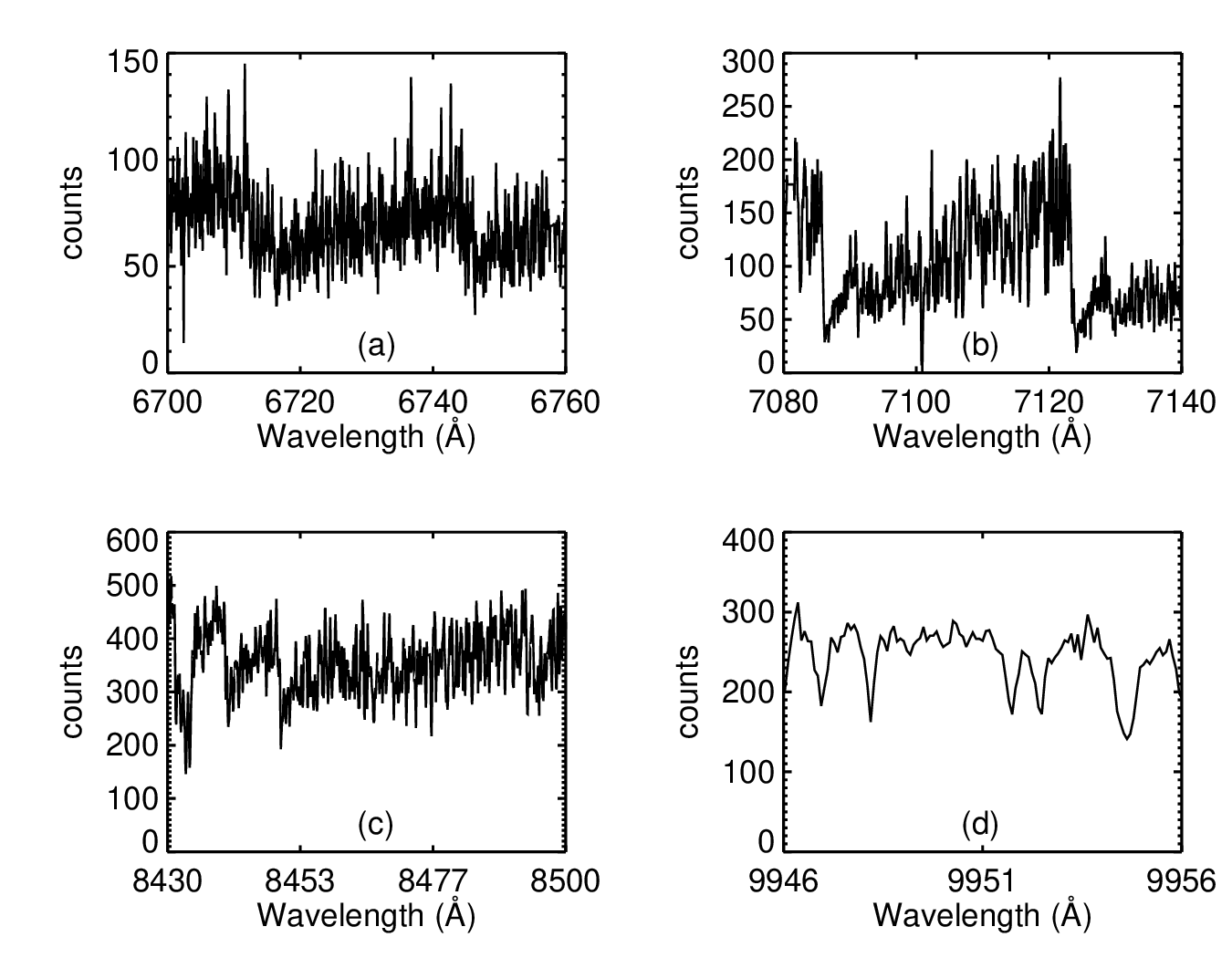}
\caption{Keck spectra of SDSS094738.45+371016.5 in the four spectral
  regions used for cross-correlation: (a) 6700-6760 \AA, (b) 7080-7140
  \AA, (c) 8430-8500 \AA, and  (d) 9946-9956 \AA.}
\label{figure:regions}
\end{figure*}

We used Gl 406 (M6) as our template because it has a similar spectral
type to our sample as well as a small (non detectable) rotation
velocity \citep{D98, MB03}.  Four wavelength regions with strong
molecular bands were used to compute independent cross-correlation
functions: 6700-6760 \AA, 7080-7140 \AA, 8430-8500 \AA~\citep[adapted
from][]{BBMW08}, and 9946-9956 \AA~\citep{Reiners07}.  Figure
\ref{figure:regions} shows an example of all four regions from one of the
stars in our sample (SDSS094738.45+371016.5). The S/N ranges from 10-60 in all
of the spectral regions for all of the stars. The three bluer spectral windows
yielded good cross-correlation functions that were consistent with
each other for all 14 program stars. In contrast, only a few good
solutions were produced using the 9946-9956\AA~window. The problem
with this spectral window may be due to the mix of broad and narrow
FeH features in the 9946-9956\AA~region; this region is more amenable
to a spectral fitting approach to measure rotational broadening
\citep[e.g.\,][]{Reiners07}.  Gl 406 is an active M6 dwarf with a
strong magnetic field \citep{RB07}, and the FeH region is also
sensitive to magnetic effects. To alleviate this problem, we followed
the same cross-correlation procedure but instead used one of our non
rotating (based on the other 3 spectral regions) program stars,
SDSS125855.13+052034.7 to cross correlate with the sample spectra.
Using this template, we were able to get better cross-correlation
functions for the 9946-9956\AA~region.

\begin{figure}
\plotone{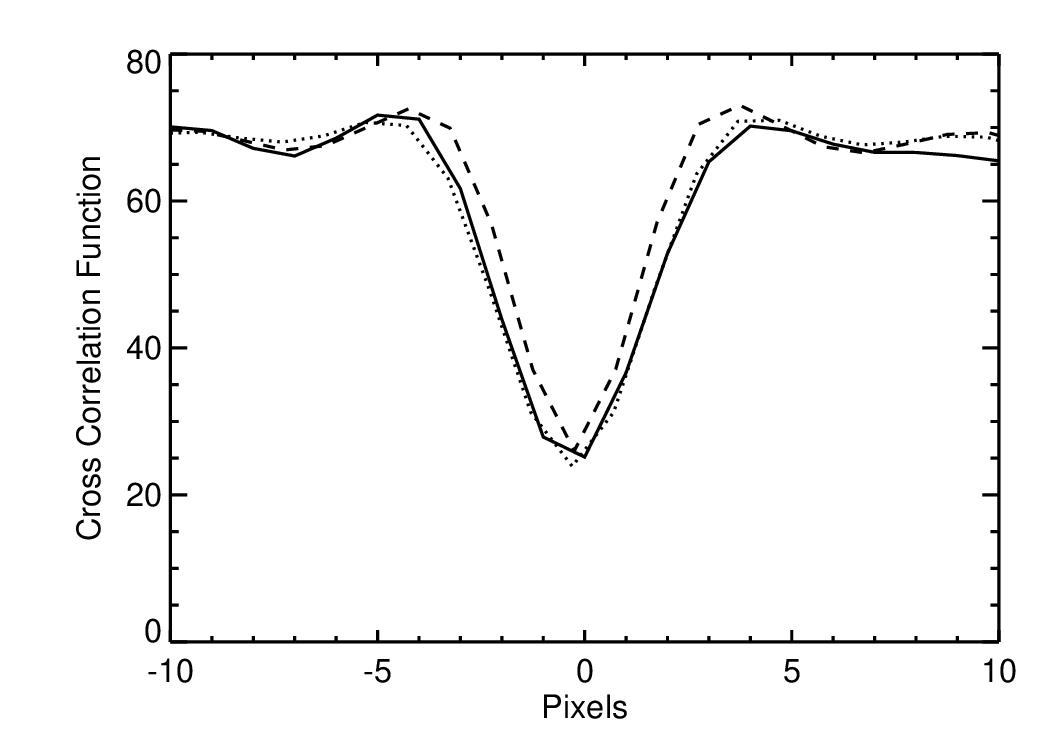}
\caption{Cross correlation function of SDSS094738.45+371016.5 with
  Gl406 in the 7080-7140 \AA~ region (solid) compared with the
  cross-correlation function of Gl406 with the best-fit rotationally
  broadened Gl406 spectrum (dotted; 6 km\,s$^{-1}$), and the
  auto-correlation function of Gl406 (dashed; 0 km\,s$^{-1}$
  broadening). The cross-correlation reveals that
  SDSS094738.45+371016.5 appears to be rotating with a velocity $\geq$
  6 km\,s$^{-1}$ despite not having any signs of activity in either the
  SDSS or Keck spectra.}
\label{figure:cc} 
\end{figure}

To measure the rotational broadening of each spectrum, we compared the
resulting cross-correlation function to that of a rotationally
broadened template. The Gl 406 and SDSS125855.13+052034.7 templates
were rotationally broadened to larger rotation velocities using the
technique of \citet{Gray92} and cross-correlated with the original
(unbroadened) template in 0.5 km\,s$^{-1}$ intervals.  The resulting spun-up cross-correlation
functions were compared with the cross-correlation functions of the
sample spectra and a $v$sin$i$ was determined based on the best fit
spun-up template.  Figure \ref{figure:cc} shows the cross-correlation
function of SDSS094738.45+371016.5 with Gl406 in the 7080-7140\AA~region (solid) compared with the cross-correlation function of Gl406
with the best-fit rotationally broadened Gl406 spectrum (dotted; 6
km\,s$^{-1}$), and the auto-correlation function of Gl406 (dashed; 0
km\,s$^{-1}$ broadening).  The cross-correlation reveals that
SDSS094738.45+371016.5 appears to be rotating with a velocity $\geq$ 6
km\,s$^{-1}$.

No detectable difference was seen between the non-rotating template and the spun-up
templates with V$_{rot}\leq$3.5 km\,s$^{-1}$ for the S/N of our typical
spectra.  We therefore used 4 km\,s$^{-1}$ as our threshold for
detectable rotation in this paper.  The $v$sin$i$ values from each
spectral window were averaged and the resulting value for each star is
reported in Table \ref{table:colors}.  For stars with multiple
spectra, the $v$sin$i$ values in Table \ref{table:colors} represent
the average of all spectral windows for all observations.  The
$v$sin$i$ variation amongst spectral windows and multiple observations
was always within the 0.5 km\,s$^{-1}$ bin size. We therefore used 0.5
km\,s$^{-1}$ as the uncertainty in the $v$sin$i$ measurements.

\subsection{Spectral Analysis}
\subsubsection{Activity}

\begin{figure*}
\plotone{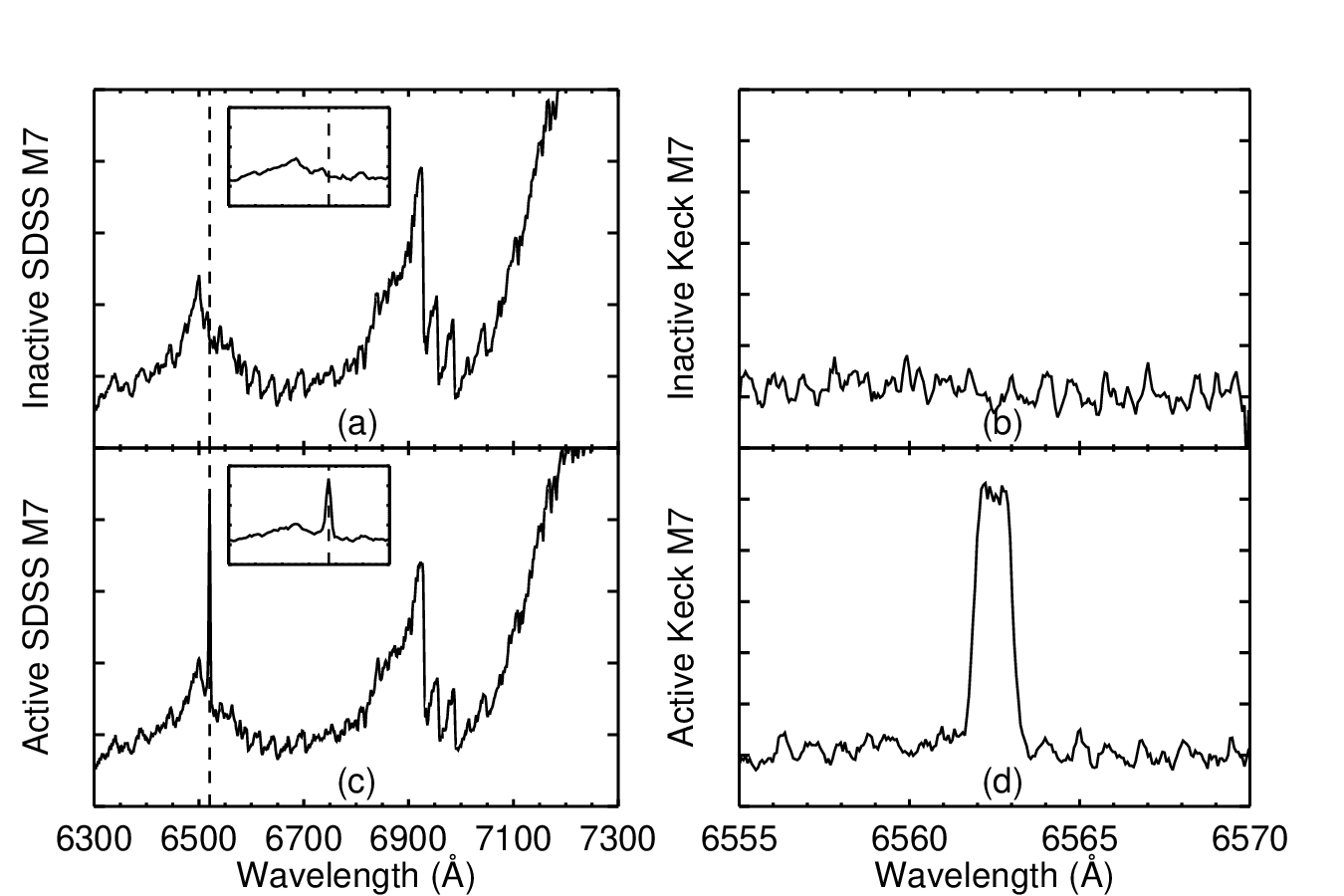}
\caption{SDSS and Keck spectra of the inactive M7 dwarf
  SDSS094738.45+371016.5 (a and b respectively) and the active M7
  dwarf SDSS162718.20+353835.7 (c and d).  Dashed lines in the SDSS
  spectra (a and c) indicate the expected position of the H$\alpha$
  emission line.  The inset panels in the SDSS spectra are zoomed in
  on the H$\alpha$ region (6500-6600 \AA). Despite their high quality,
  it is clear that no H$\alpha$ emission is detected in either of the
  inactive spectra.}
\label{figure:inact}
\end{figure*}

All of the spectra were spectral typed by eye using the Hammer
spectral analysis package \citep{Covey07} on the SDSS spectra.  We
measured the equivalent widths (EW) of the H$\alpha$ emission lines in
both the SDSS and Keck spectra.  The Keck spectra are more sensitive
to low levels of emission; they can be distinguished better against
the pervasive molecular features. Our EWs are really psuedo-EWs, since
there is no real continuum in the spectra.  This has different effects
at low and high resolution, which is one reason that EWs tend to be
smaller when measured from high resolution spectra.  Almost all of our
targets chosen to be inactive at low resolution proved inactive even
at high resolution, and the EWs when detected were similar. In one
case (SDSS083231.52+474807.7), the high resolution spectrum seems to
have caught a small flare.  Figure \ref{figure:inact} shows the SDSS and Keck
spectra of the inactive M7 dwarf SDSS094738.45+371016.5 (a and b
respectively) and the active M7 dwarf SDSS162718.20+353835.7 (c and d).
Dashed lines in the SDSS spectra (a and c) indicate the expected
position of the H$\alpha$ emission line.  The inset panels in the SDSS
spectra are zoomed in on the H$\alpha$ region (6500-6600 \AA). Despite
their high quality, it is
clear that no H$\alpha$ emission is detected in either of the inactive spectra.

\begin{deluxetable*}{lccrrrrrrcccc}
\tablewidth{0pt}
\tablewidth{0pt}
\tablecolumns{13} 
\tabletypesize{\scriptsize}
\tablecaption{Measured Attributes}
\renewcommand{\arraystretch}{.6}
\tablehead{
\colhead{Name}&
\colhead{Sp. Type}&
\colhead{$v$sin$i$}&
\multicolumn{6}{c}{Equivalent Width (\AA)}&
\colhead{TiO5\tablenotemark{c}}&
\colhead{CaH2\tablenotemark{c}}&
\colhead{CaH3\tablenotemark{c}}&
\colhead{Rossby}\\
\colhead{ }&
\colhead{ }&
\colhead{(km\,s$^{-1}$)}&
\colhead{H$\alpha$\tablenotemark{a}}&
\colhead{H$\alpha$\tablenotemark{b}}&
\colhead{H$\beta$\tablenotemark{a}}&
\colhead{H$\gamma$\tablenotemark{a}}&
\colhead{H$\delta$\tablenotemark{a}}&
\colhead{CaII K\tablenotemark{a}}&
\colhead{ }&
\colhead{ }&
\colhead{ }&
\colhead{Num.}}
\startdata
SDSS011012.22$-$085627.5 &	M7 &	$<$3.5 &	$<$0.2 &$<$0.2 &  $<$0.6
&    $<$1.7 &     $<$4.3 &    $<$0.9 &	0.17 & 0.27 & 0.56 &\nodata\\  
SDSS021749.99$-$084409.4 &	M6 &	4.0$\pm$0.5 &	$<$0.1 & $<$0.1 &   $<$0.4
&    $<$2.2 &   $<$1.7 &   $<$0.8 &	0.28 & 0.35 & 0.67 & 0.03\\
SDSS023908.41$-$072429.3 &	M7 &	$<$3.5 &	$<$0.1 & 0.3 &	$<$0.8 &
$<$7.0 &    \nodata &    \nodata  &	0.23 & 0.32 & 0.61 & \nodata\\
SDSS072543.94+382511.4 &	M7 &	$<$3.5 &	$<$0.1 &$<$0.1 & $<$0.2 &
$<$0.5 &   $<$0.5 &    $<$1.1 &	0.27 & 0.35 & 0.67 & \nodata\\
SDSS083231.52+474807.7 &	M6 &	4.0$\pm$0.5 &	$<$0.1 & 3.9\tablenotemark{d} &	$<$0.2 &
$<$0.5 &     $<$0.5 &     1.1 &	0.28 & 0.34 & 0.64 & 0.03\\
SDSS094720.07$-$002009.5 &	M7 &	6.5$\pm$0.5 &	3.4 &	4.0 &	4.2 &
17.2 &     2.6 &    12.2 &	0.22 & 0.28 & 0.57 & 0.01\\
SDSS094738.45+371016.5 &	M7 &	6.0$\pm$0.5 &	$<$0.2 &$<$0.1 & $<$0.2 &
$<$0.9 &    $<$1.8 &   $<$1.0 &	0.19 & 0.27 & 0.59 & 0.01\\
SDSS110153.86+341017.1 &	M7 &	$<$3.5 &	$<$0.2 &$<$0.1 & $<$0.2 &
$<$1.1 &   $<$1.1 & $<$0.8 &	0.22 & 0.29 & 0.60 & \nodata\\
SDSS112036.08+072012.7 &	M7 &	$<$3.5 &	$<$0.1 &$<$0.1 & $<$0.2 &
$<$0.3 &     $<$0.4 &    $<$0.5 &	0.26 & 0.32 & 0.66 & \nodata\\
SDSS125855.13+052034.7 &	M7 &	$<$3.5 &	0.9 &	$<$0.1 &	1.8 &
7.2 &     4.5 &     5.8 &	0.25 & 0.31 & 0.62 & \nodata\\
SDSS151727.72+335702.4 &	M7 &	4.5$\pm$0.5 &	$<$0.2 &$<$0.1 & $<$0.2 &
$<$0.6 &    $<$0.8 &   $<$0.3 &	0.26 & 0.35 & 0.70 & 0.02\\
SDSS162718.20+353835.7 &	M7 &	8.0$\pm$0.5 &	5.9 &	5.4 &	19.7 &
37.1 &    32.4 &    26.5 &	0.24 & 0.20 & 0.61  & 0.01\\
SDSS220334.10+130839.8 &	M7 &	$<$3.5 &	$<$0.1 &$<$0.1 & $<$0.6 &
$<$2.8 &  $<$2.3 &   $<$2.3 &	0.22 & 0.32 & 0.66 & \nodata\\
SDSS225228.50$-$101910.9 &	M7 &	$<$3.5 &	$<$0.2 &$<$0.2 & $<$0.7 &
$<$9.8 & \nodata & \nodata &	0.19 & 0.26 & 0.59 & \nodata\\
\enddata
\tablenotetext{a}{Equivalent width or 3$\sigma$ upper limit measured from the SDSS spectrum}
\tablenotetext{b}{Equivalent width or 3$\sigma$ upper limit measured from the Keck spectrum}
\tablenotetext{c}{Bandhead measured from the SDSS spectrum}
\tablenotetext{d}{Keck spectrum observerd during a small flare}
\label{table:colors}
\end{deluxetable*}

We also measured the H$\beta$,
H$\gamma$, H$\delta$, and CaII K emission lines from the SDSS spectra.
The EW values are presented in Table \ref{table:colors}.  For the stars in which no lines were detected, we computed EW upper limits based on the required EW needed to detect an emission line at the 3$\sigma$ confidence level. These are also reported in Table \ref{table:colors}. Using the
relations from \citet[H$\alpha$]{WHW04} and \citet[other lines]{WH08}, we converted the EWs and upper limits for all of the lines
to fractions of the bolometric luminosity
(e.g. L$_{\rm{H\alpha}}$/L$_{\rm{bol}}$, etc.).  The
L$_{\rm{line}}$/L$_{\rm{bol}}$ values are presented in Table
\ref{table:colors2}.

We estimated the Rossby number ($R_0$=Period/$\tau_{c}$) for the 6
stars with measured rotation velocities, following the prescription of
\citet[$\tau_c$=70 days]{RBB08}, and assuming stellar radii of 0.12
R$_{\odot}$ and 0.1 R$_{\odot}$ for M6 and M7 dwarfs respectivley
\citep{NLDS}.  Our estimated $R_0$ values are included in Table
\ref{table:colors}. All of the estimated Rossby numbers are in the
regime of activity ``saturation'' ($R_0<0.1$) described in
\citet{RBB08}.

\subsubsection{Metallicity}

We measured the TiO5, CaH2, and CaH3 molecular band indices
\citep{Reid95} from the SDSS spectra, features that probe the relative
metallicities of M dwarfs \citep{Lepine03, Burgasser06, W08}. The
combination of these indices (CaH2+CaH3 vs. TiO5) has been used to
separate the dwarf population ([Fe/H]$\sim$0) from the subdwarf
population ([Fe/H]$\sim$-1). We include the index value for each
molecular band in Table \ref{table:colors}.  All of the index values
are consistent with disk dwarf metallicity (near Solar).

\subsubsection{Dynamics}

We assigned a Galactic population to each star using the 3-D
space motions computed in \citet{W08} and the criteria of
\citet{Leggett92}, where young disk stars (YD) have -20 $<$ U $<$ 50,
-30 $<$ V $<$ 0, and -25 $<$ W $<$ 10. Stars outside this ellipse or
with $|$W$|$ $>$ 50 are old disk (OD) and stars near the edge but with
$|$W$|$ $<$ 50 are classified as young-old disk stars (YOD).  All of
the dynamic classifications are included in Table
\ref{table:colors}. While our sample includes all 3 of the components,
it is dominated by the old disk and yound old disk populations; 12 of
14 stars are OD or YOD.

\begin{deluxetable*}{lccccccccc}
\tablewidth{0pt}
\tablewidth{0pt}
\tablecolumns{10} 
\tabletypesize{\scriptsize}
\tablecaption{Measured Attributes II}
\renewcommand{\arraystretch}{.6}
\tablehead{
\colhead{Name}&
\colhead{Spec.}&
\colhead{$v$sin$i$}&
\colhead{L$_{\rm{H}\alpha}$/L$_{bol}$\tablenotemark{a}}&
\colhead{L$_{\rm{H}\alpha}$/L$_{bol}$\tablenotemark{b}}&
\colhead{L$_{\rm{H}\beta}$/L$_{bol}$\tablenotemark{a}}&
\colhead{L$_{\rm{H}\gamma}$/L$_{bol}$\tablenotemark{a}}&
\colhead{L$_{\rm{H}\delta}$/L$_{bol}$\tablenotemark{a}}&
\colhead{L$_{\rm{CaII K}}$/L$_{bol}$\tablenotemark{a}}&
\colhead{Dyn.\tablenotemark{c}}\\
\colhead{}&
\colhead{Type}&
\colhead{(km\,s$^{-1}$)}&
\colhead{($\times10^{-6}$)}&
\colhead{($\times10^{-6}$)}&
\colhead{($\times10^{-6}$)}&
\colhead{($\times10^{-6}$)}&
\colhead{($\times10^{-6}$)}&
\colhead{($\times10^{-6}$)}&
\colhead{Pop.}}
\startdata
SDSS011012.22$-$085627.5 &	M7 &   $<$3.5 &	     $<$1.0 & $<$1.0 &   $<$0.5 &     $<$0.3   &   $<$0.9  &    $<$0.2  & YOD\\
SDSS021749.99$-$084409.4 &	M6 &	4.0$\pm$0.5 &	     $<$2.0 & $<$2.0 &   $<$1.4  &    $<$2.4   &   $<$1.5  &    $<$0.8  & YD\\
SDSS023908.41$-$072429.3 &	M7 &	$<$3.5 &	     $<$0.5 & 1.6  &     $<$0.6     & $<$1.4  &    \nodata   &  \nodata & YOD\\
SDSS072543.94+382511.4 &	M7 &	$<$3.5 &	     $<$0.5 & $<$0.5 &   $<$0.2  &    $<$0.1   &   $<$0.1  &    $<$0.2 & OD \\
SDSS083231.52+474807.7 &	M6 &	4.0$\pm$0.5 &	     $<$2.0 & 69.0  &    $<$0.7   &   $<$0.6   &   $<$0.5 & 1.1 & OD\\
SDSS094720.07$-$002009.5 &	M7 &	6.5$\pm$0.5 &	     18.0 & 21.0 & 3.4 & 3.4 & 0.52 & 2.4 & YOD\\
SDSS094738.45+371016.5 &	M7 &	6.0$\pm$0.5 &	     $<$1.0 & $<$0.5 &   $<$0.2  &    $<$0.2   &   $<$0.4  &    $<$0.2 & OD\\
SDSS110153.86+341017.1 &	M7 &	$<$3.5 &	     $<$1.0 & $<$0.5 &   $<$0.2  &    $<$0.2   &   $<$0.2  &    $<$0.2 & OD\\
SDSS112036.08+072012.7 &	M7 &	$<$3.5 &	     $<$0.5 & $<$0.5 &   $<$0.2  &    $<$0.1   &   $<$0.1  &    $<$0.1 & OD\\
SDSS125855.13+052034.7 &	M7 &	$<$3.5 &	   4.68   &   $<$0.5 & 1.4 & 1.4 & 0.90 & 1.2 & OD \\
SDSS151727.72+335702.4 &	M7 &	4.5$\pm$0.5 &	      $<$1.0 & $<$0.5 &  $<$0.2  &    $<$0.1   &   $<$0.2  &    $<$0.1 & YOD\\
SDSS162718.20+353835.7 &	M7 &	8.0$\pm$0.5 &	     31.0 & 28.0 & 16.0 & 7.4 & 6.5 & 5.3 & YOD  \\
SDSS220334.10+130839.8 &	M7 &	$<$3.5 &	      $<$0.5 & $<$0.5 &  $<$0.5  &    $<$0.6   &   $<$0.5  &    $<$0.4 & YD\\
SDSS225228.50$-$101910.9 &	M7 &	$<$3.5 &	      $<$1.0 & $<$1.0 &  $<$0.6  &    $<$2.0   &   \nodata  &   \nodata & YOD\\
\enddata
\tablecomments{The L$_{line}$/L$_{\rm{bol}}$ values were calculated using the $\chi$
  conversion factors derived by \citet{WHW04} and \citet{WH08}.}
\tablenotetext{a}{Equivalent width or 3$\sigma$ upper limit measured from the SDSS spectrum}
\tablenotetext{b}{Equivalent width or 3$\sigma$ upper limit measured from the Keck spectrum}
\tablenotetext{c}{Dynamical Populations based on the criteria in
  \citet{Leggett92}}
\label{table:colors2}
\end{deluxetable*}

\section{Observational Results}

Six of the fourteen M6-M7 dwarfs in our sample have detectable
rotation. Three of the rotating stars have measurable activity but the
other 3 show no signs of activity in any of the emission lines in
either the SDSS or Keck spectra.  The cross-correlation shown in
Figure \ref{figure:cc} (SDSS094738.45+371016.5) is an example of one
of the inactive M7 dwarfs that appears to be rotating despite not
being magnetically active.

\begin{figure*}
\plotone{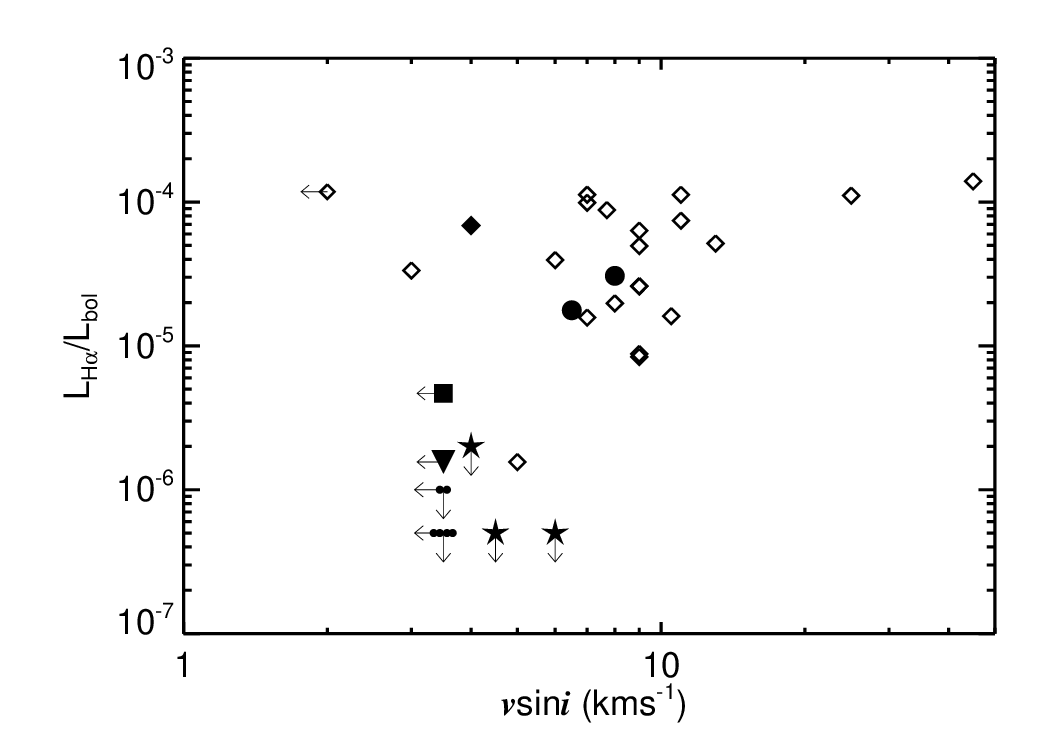}
\caption{L$_{\rm{H}\alpha}$/L$_{\rm{bol}}$ (activity) as a function of
  $v$sin$i$ (rotation) for the M6-M7.5 dwarfs from this paper (filled
  symbols) and previous studies \citep[open diamonds]{D98, MB03,
    Reiners07}.  L$_{\rm{H}\alpha}$/L$_{bol}$ values were calculated
  from equivalent width measurements using the $\chi$ conversions of
  \citet{WHW04}.  Our sample includes M6-M7 dwarfs with both measured
  rotation as well as activity from the SDSS and Keck spectra (filled
  circles), measured rotation and activity from the Keck spectra (CaII
  K detected in SDSS; filled diamonds), activity from the Keck spectra
  but no rotation (filled triangles), activity from the SDSS spectra
  but no rotation (filled squares), no rotation or activity
  (horizontally aligned dots) and measured rotation but no activity
  (filled stars). Upper limits in both velocity and activity denote
  the levels to which our sample (and previous studies) could probe.
  All previous M6-M7.5 dwarfs were found to be active, while 9 of the
  14 stars in our sample show no activity in either the SDSS or Keck
  spectra.  The dearth of inactive late-type M dwarfs in previous
  studies is due to a selection effect that biases nearby samples to
  younger, more active stars \citep{W06,W08}. 3 of our stars show
  strong evidence for rotation despite having no activity. Their
  $v$sin$i$ velocities are well above the 0.7 km\,s$^{-1}$ required to
have $R_0 > 0.1$ and be beyond the ``saturated'' regime \citep{RBB08}.}
\label{figure:vsini} 
\end{figure*}

Figure \ref{figure:vsini} shows L$_{\rm{H}\alpha}$/L$_{\rm{bol}}$
(activity) as a function of $v$sin$i$ for the M6-M7.5 dwarfs from this
paper (filled symbols) and previous studies \citep[open
diamonds]{D98, MB03, Reiners07}.  L$_{\rm{H}\alpha}$/L$_{\rm{bol}}$ values
were calculated from EW measurements using the $\chi$ conversions of
\citet{WHW04}.  Our sample includes M6-M7 dwarfs with both measured
rotation as well as activity from the SDSS and Keck spectra (filled
circles), measured rotation and activity from the Keck spectra (CaII K
detected in SDSS; filled diamonds), activity from the Keck spectra but
no rotation (filled triangles), activity from the SDSS spectra but no
rotation (filled squares), no rotation or activity (horizontally
aligned dots) and measured rotation
but no activity (filled stars). Upper limits in both velocity and
activity denote the levels to which our sample (and previous studies)
could probe.  All previous M6-M7.5 dwarfs were found to be active,
while 9 of the 14 stars in our sample show no activity in either the
SDSS or Keck spectra.  The lack of activity is not surprising since
that was our main selection criterion, however it highlights the fact
that we are probing a sample with very different properties than
previously studied.

The two most active stars in our sample are also the fastest rotators,
and 6 of the inactive stars have no detectable rotation.  Two of the
active stars show no sign of rotation, which may be due to inclination
effects (although the incidence in that case would be higher than
statistically expected, if such small numbers can be said to have any
significance).\footnote{Using the statistical formalism outlined in
  \cite{BBMW08}, the probability of selecting 2 out of 5 active stars with
  the inclination required to alter their observed $v$sin$i$
  velocities from the mean detected velocity of 6 km\,s$^{-1}$ to 3.5 km\,s$^{-1}$
  (non-detection) is 24\%.}  A similar excess of active
non-rotators was found by \citet{BBMW08} at spectral types M3-M5.  All
6 of the stars with detectable rotation have estimated Rossby numbers
that are beyond (smaller than, by an order of magnitude) the activity
``saturation'' threshold of $R_0$=0.1 discussed in \citet[the stars would need
to have V$_{rot}<$0.7 km\,s$^{-1}$ to have $R_0>$0.1]{RBB08}.
However, none of these rotators show activity at the ``saturated''
level of L$_{\rm{H}\alpha}$/L$_{\rm{bol}}\sim10^{-4}$.  Most intriguing are
the three M7 dwarfs that show no activity but are clearly detected
rotators.  We discuss this in more detail in \S5.

Neither the metallicity nor dynamic population analyses reveal any
significant trends with either rotation velocity or activity.  This is
not surprising, since both measurements are only relevant in a
statistical sense.  However, all but one of the stars with a detected
rotation comes from the YD or YOD population, and the one rotating OD shows no
signs of activity (as do one of each from the other two populations).

\section{Discussion}

We conducted high resolution spectral observations of 14 M6-M7 dwarfs
and found $v$sin$i$ rotation velocities for 6 for the stars.  Three of
the stars showed both activity and rotation, 6 of the stars showed
neither rotation nor activity, 2 of the stars showed activity but no
rotation and 3 stars showed rotation but no activity.  These results
are in contrast with previous studies that found a strong connection
between rotation and activity in all (active) M6-M7 dwarfs
\citep{D98,MB03, Reiners07}.  As can be seen in Figure
\ref{figure:vsini}, those studies produced one out of 20 stars which
was active but not rotationally broadened (very reasonably explained
by the inclination effect; there is a 94\% chance of selecting one out
of 20 with the inclination needed to take it from 4 km\,s$^{-1}$ to 2
km\,s$^{-1}$), and one which showed rotation and low (but not zero)
activity.  Our sample is the first rotation study to include by design
M6-M7 dwarfs that are inactive.

Whereas previous studies sampled a range of dynamic populations and
therefore stellar ages \citep{D98, MB03}, our sample was selected from
a more distant, mostly inactive and potentially older population
\citep[M7 dwarfs have activity lifetimes of $\sim$8 Gyr;][]{W08}.
Although our sample is not representative of the entire
M6-M7 population, it does examine a different part of parameter space
than previously sampled and observes something not previously seen:
rotation in inactive M7 dwarfs.  Currently, there is no clear
explanation for the rotation-activity patterns seen in our sample.
However, we present a number of possible explanations that may account
for this unusual dataset.

One possible explanation is that rotation and activity are not
strongly linked in the fully convective envelopes of M6-M7
dwarfs. Previous studies only examined active M6-M7 dwarfs, many of
which have activity levels with no notable correlation with rotation
beyond a general tendancy for rotators to lie in a broadly spread
``saturated'' activity regime.  In fact, several active dwarfs have no
detectable rotation (see Figure \ref{figure:vsini}).  Besides
saturation, another explanation for the lack of a strong relationship
between activity and rotation is inclination; the rotation axes of the
stars may be inclined and therefore have reduced line-of-sight
components to their rotation velocities.  On the other hand, the 3 inactive stars with
detected rotations certainly cannot be resolved by simply imposing
inclination effects, which can only serve to make their true rotations
even higher.

Another possibility is that there is a rotationally-induced threshold
for activity at V$_{rot}>$6 km\,s$^{-1}$ \citep{Hawley96}. This would
explain the inactive rotators (they are below the threshold) and
require that active stars with $v$sin$i$ rotation velocities less than
the threshold have inclined rotation axes. Combining all of the
previous rotation studies for M6-M7 (Figure \ref{figure:vsini}), we
see that 7 of the 26 active dwarfs have $v$sin$i\leq$ 6 km\,s$^{-1}$;
the rest of the stars are rotating several km\,s$^{-1}$ above the
detection thresholds. To quantify this possibility, we follow the
statistical analysis of \citet{BBMW08}, where we compute the
probability (from the binomial distribution) of observing n objects
with inclination $i$ $<$ $i_{crit}$.  We find that the probability
of randomly drawing a sample of 7 stars with the inclinations
necessary to take the slow but active stars from a V$_{rot}$=6.5
km\,s$^{-1}$ to the observed $v$sin$i$ is 85.7\%.  If we instead
calculate the more conservative probability of drawing inclinations
that take the 7 stars from 8 km\,s$^{-1}$ (the median $v$sin$i$) to
their respective $v$sin$i$ values, we obtain 55.9\%.  In this
scenario, the inactive stars, which are presumably older, have spun
down enough to fall below the rotation-activity threshold.

In earlier M dwarfs, the mere detection of rotation implies that the
star is in the ``saturation'' regime; an effect that should have
become even more pronounced for these even smaller objects (assuming
the convective overturn time is similar).  All of our detected
rotators have estimated Rossby numbers that are beyond the
``saturation'' threshold for early-type M dwarfs \citep{RBB08}.  One
possibility is that in these fully convective late-type M dwarfs, the
Rossby number ``saturation'' threshold might be substantially smaller.
However, this model would imply that stars go from fully active to
completely inactive over a small range of Rossby numbers - something
that contradicts what is seen in early-type M dwarfs, where the
activity falls smoothly with increasing Rossby number \citep{Kiraga07,
  RBB08}. It is clear that the velocity thresholds proposed here would
have to be lowered (as a function of spectral type) to be compatible
with the results of \citet{RBB08}; too many of their stars are active
below $v$sin$i=$ 6 km\,s$^{-1}$ (including several in the M5-6 spectral
range).

% It could be that the decreasing convective flux causes something like that, although I would have thought it would go the other way (rotation having an easier time forcing the dynamo to a saturated state with weak convection).

Of course one additional possibility is that M6-M7 dwarfs might have
activity cycles and that we are observing the inactive stars at a
minimum in their cycles.  Although this might be observed for a small
number of stars, it is unlikely that an activity minimum can explain
the majority of inactive M dwarfs seen in the Milky Way.  \citet{W08}
showed that the activity fractions of M dwarfs are strongly correlated
with the height above the Galactic plane (and therefore age), a trend
that cannont be explained by an activity cycle.

The dearth of a large sample of nearby, inactive late-type dwarfs
precludes large-scale follow-up that would further test the
age-rotation-activity relation of fully convective low-mass dwarfs.
However, future efforts to extend our analysis to both inactive M5 and
M8 dwarfs will help elucidate the problem.  The multi-epoch
photometric observations produced by all sky time-resolved surveys
(e.g. Pan-STARRS and LSST) will produce rotation periods for a
large number of late-type M dwarfs that will eliminate the inclination
effects.  Of course, inactive M dwarfs present a problem to this
method in that they may not exhibit photometric variability (if they
don't produce starspots).  The combination of these and other future
observations will help advance the understanding of what role rotation
plays in the production of magnetic fields in fully convective objects, and provide crucial
constraints for theoretical models of fully convective dynamos.  They
will also be a key component to understanding the feedback between
magnetic fields and stellar angular momentum evolution through stellar
activity.

\section{Acknowledgments}
The authors would like to thank Matt Browning, Ansgar Reiners, Suzanne
Hawley, Lucianne Walkowicz, Kevin Covey, Adam Burgasser and John
Bochanski for useful discussions while conducting this study.

G.B. acknowledges support from the NSF through grant AST-0606748.
Some of the data presented herein were obtained at the
W.M. Keck Observatory, which is operated as a scientific partnership among the
California Institute of Technology, the University of California and
the National Aeronautics and Space Administration. The Observatory was
made possible by the generous financial support of the W.M. Keck
Foundation.

Funding for the Sloan Digital Sky Survey (SDSS) and SDSS-II has been
provided by the Alfred P. Sloan Foundation, the Participating
Institutions, the National Science Foundation, the U.S. Department of
Energy, the National Aeronautics and Space Administration, the
Japanese Monbukagakusho, and the Max Planck Society, and the Higher
Education Funding Council for England. The SDSS Web site is
http://www.sdss.org/.

The SDSS is managed by the Astrophysical Research Consortium (ARC) for
the Participating Institutions. The Participating Institutions are the
American Museum of Natural History, Astrophysical Institute Potsdam,
University of Basel, University of Cambridge, Case Western Reserve
University, The University of Chicago, Drexel University, Fermilab,
the Institute for Advanced Study, the Japan Participation Group, The
Johns Hopkins University, the Joint Institute for Nuclear
Astrophysics, the Kavli Institute for Particle Astrophysics and
Cosmology, the Korean Scientist Group, the Chinese Academy of Sciences
(LAMOST), Los Alamos National Laboratory, the Max-Planck-Institute for
Astronomy (MPIA), the Max-Planck-Institute for Astrophysics (MPA), New
Mexico State University, Ohio State University, University of
Pittsburgh, University of Portsmouth, Princeton University, the United
States Naval Observatory, and the University of Washington.

\bibliographystyle{apj}

%\begin{thebibliography}
%\nocite{*} 
%\bibliography{ms_101808}

%\end{thebibliography}

%table

%include kinematic ages

\end{document}